%% file: main.tex
\documentclass[a4paper,11pt]{article}
\usepackage{pos}
\usepackage[capitalize]{cleveref}
\usepackage{subcaption}
\usepackage{enumitem}
\usepackage{wrapfig}       
\usepackage[left]{lineno}

\title{Calibration and Performance of the Surface Scintillator Detector of the Pierre Auger Observatory}
\ShortTitle{Surface Scintillator Detector Calibration and Performance}

\author*[ab]{Matteo Conte}

\affiliation[a]{INFN Sezione di Lecce,  Lecce, Italy}
\affiliation[b]{Università del Salento, Lecce, Italy}

\onbehalf{for the Pierre Auger Collaboration$^c$}
\affiliation[c]{Observatorio Pierre Auger, Av.\ San Mart{\'\i}n Norte 304, 5613 Malarg\"ue, Argentina\\
Full author list: {\rm\url{https://www.auger.org/archive/authors_icrc_2025.html}}}

\emailAdd{spokespersons@auger.org}

\abstract{The Pierre Auger Observatory has led to significant advances in our understanding of ultra-high-energy cosmic rays.
These new insights have driven a major upgrade of the Observatory, known as AugerPrime, through which the experiment has entered its Phase\,II, a new period of data collection.
A key part of the upgrade is adding surface scintillator detectors (SSD) on top of the existing water-Cherenkov detectors (WCD).
The main goal is to leverage their different responses to the electromagnetic and muonic shower components, enhancing the reconstruction of the primary cosmic-ray mass.
In this contribution, we present the methods that involve analyzing peak and charge distributions of atmospheric muons for accurate calibration during extensive air-shower event reconstruction, along with the development
of a rate-based algorithm for independent calibration.
We also show the performance of the SSDs with Phase-II data, including PMT reliability
and stability of key parameters, such as gain and signal-to-noise ratio.}

\ConferenceLogo{icrc_logo}

\FullConference{%
39th International Cosmic Ray Conference (ICRC2025)\\
15 -- 24 July, 2025\\
Geneva, Switzerland}

\begin{document}
\maketitle

\section{Introduction}

The Pierre Auger Observatory, located in Malargüe, Argentina, is the world's largest cosmic ray detector, combining a surface array of over 1600 Water-Cherenkov Detectors (WCDs) with 27 fluorescence telescopes.
Its hybrid design enables precise measurements of extensive air showers from ultra-high-energy cosmic rays (UHECRs).
Over the years, the Observatory has produced key results on the energy spectrum, anisotropy, and mass composition of cosmic rays.

To enhance its capabilities, the Observatory has undergone a major upgrade -- AugerPrime -- which includes the addition of Surface Scintillator Detectors (SSDs) above each WCD.
These detectors improve the separation of electromagnetic and muonic components, enhancing mass composition studies.
In this contribution, we present the SSD calibration methods based on atmospheric muons, including a novel rate-based algorithm, and evaluate the detector performance using Phase-II data.

\section{The Surface Scintillator Detector}

\begin{figure}
    \centering
    \includegraphics[width=0.8\textwidth]{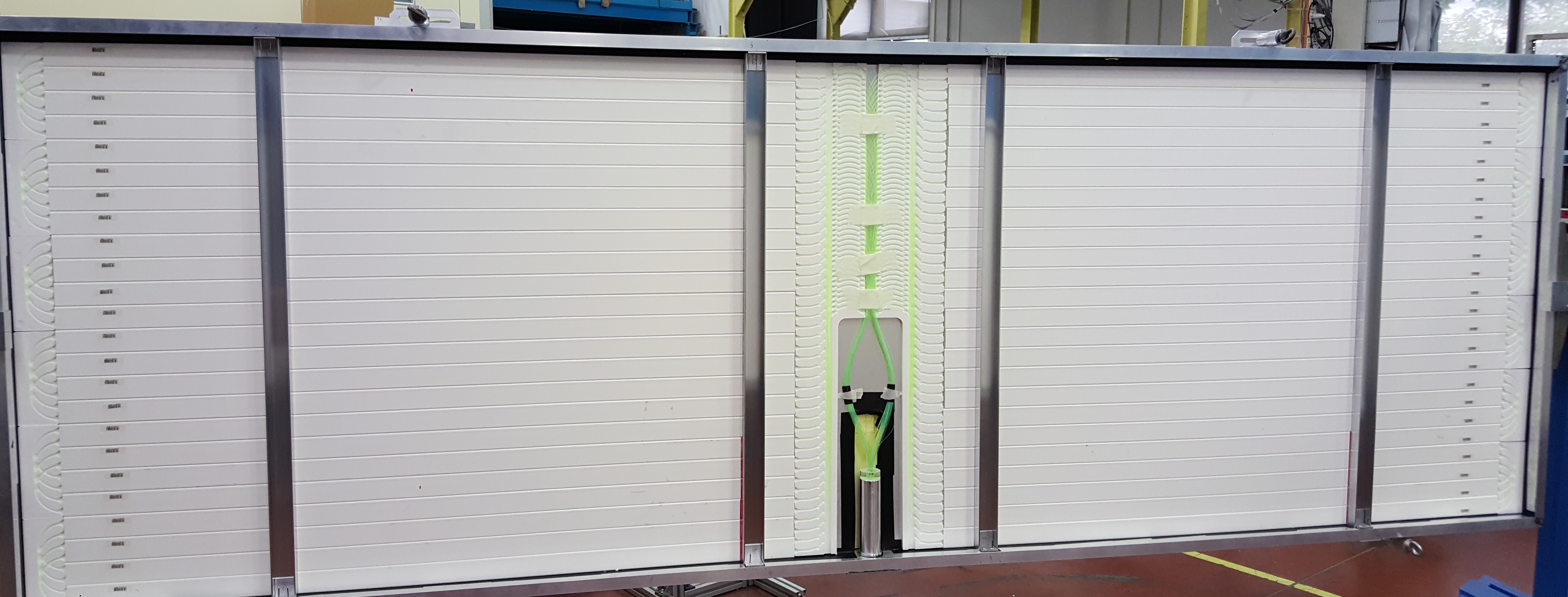}
    \caption{Photograph of an open SSD at the construction phase.}
    \label{f:ssd_schematic}
\end{figure}

The SSD module consists of two scintillator panels composed of plastic scintillator bars, encased in an aluminium box, with a photomultiplier (PMT) housed between the panels.
The total active area of the scintillators in a module is $3.84\,\text{m}^2$.
The active part of each of the scintillator panels is composed of 24 scintillator elements (bars) of
1.6\,m length, 5\,cm width, and 1\,cm thickness.
Each bar houses two horizontal holes through which wavelength-shifting (WLS) fibers are guided.
The fibers are bundled and glued with optical cement in a PMMA (poly(methyl methacrylate)) cylinder, a so-called ``cookie'' whose front window is connected to the PMT, a bi-alkali Hamamatsu R9420, 1.5-inch diameter, with 18\% quantum efficiency at a wavelength of 500\,nm \cite{SSD}.
The picture of open SSD assembled in the laboratory is shown in \cref{f:ssd_schematic}. 

The PMT is connected to the acquisition electronics, with the anode read out by a 12-bit FADC operating at 120\,MHz. 
This results in a time binning of approximately 8.33\,ns. 
To interpret the digital readings from the FADCs, it is necessary to define a unit that characterizes the energy deposited by minimum-ionizing particles (MIP).
This approach is analogous to the one already adopted for the WCD, where signals are expressed in terms of the equivalent charge produced by vertical-centered through-going muon referred to as VEM (vertical-equivalent muon).

The 3 WCD and the SSD PMTs signals are divided into four high-gain and four low-gain signals, respectively.
The gain ratio is set by the electronics design to 128 for the SSD channel and to 32 for the WCD PMTs~\cite{Fabio}.
This configuration extends the dynamic range above 20\,000\,MIP, in line with the increased dynamic range of the WCD achieved through the addition of the small PMT (see section \ref{dynamic_range}).

\section{Calibration methods}

\subsection{Histogram Based Calibration}

\begin{figure}
    \centering
    \begin{subfigure}[b]{0.4\textwidth}
        \centering
        \includegraphics[height=5cm]{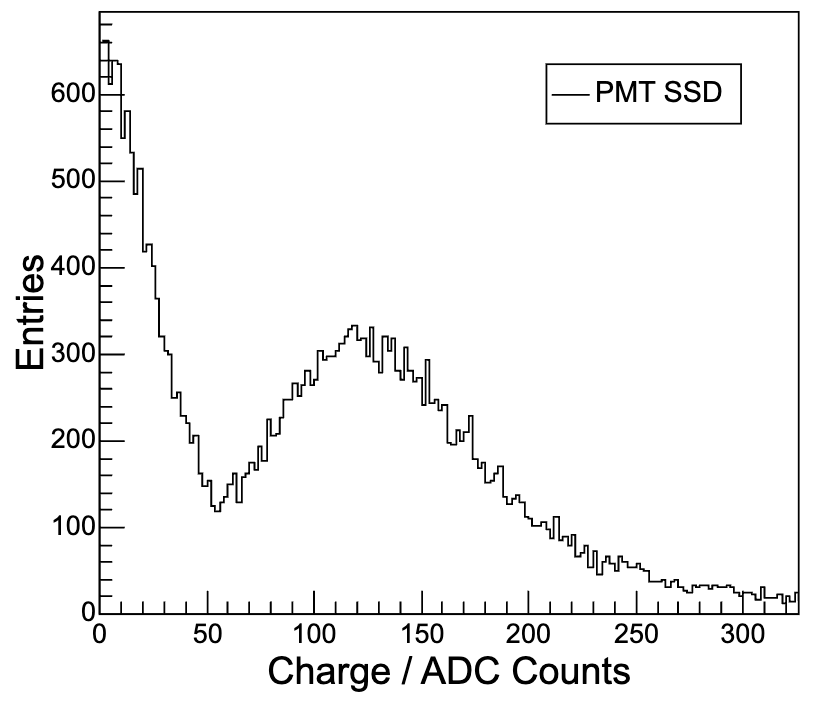}
    \end{subfigure}
    \hfill
    \begin{subfigure}[b]{0.55\textwidth}
        \centering
        \includegraphics[height=5cm]{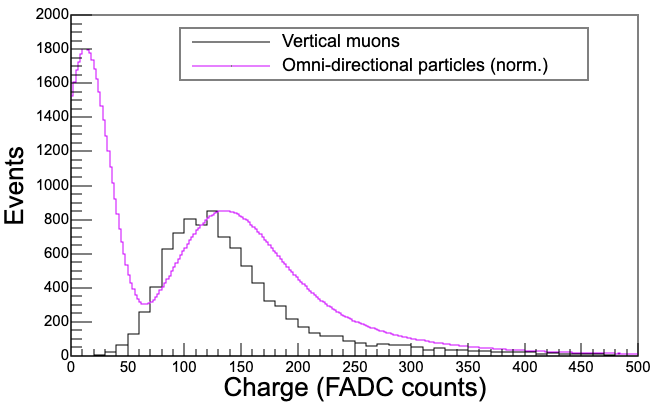}
    \end{subfigure}
    \caption{\textit{Left:} Example of calibration histograms for the PMT of one scintilltor. \textit{Right:} Estimation of the correction factor from omnidirectional to vertical muon distribution. }
    \label{f:calib_SSD}
\end{figure}

Calibration of MIP and VEM in ADC counts is based on atmospheric muons, which are abundant and energetic.
For the WCD, where vertical muons cannot be isolated, charge spectra from muons arriving at all angles are continuously collected as calibration histograms.

A similar method is used for the SSD, selecting only events where a WCD PMT detects a muon-like signal, improving sample purity by exploiting the distinct response of WCD to muons and electromagnetic particles.
Histograms are recorded every 61 seconds and used to estimate calibration units.
An example of charge calibration histogram for the SSD is shown in the left plot of \cref{f:calib_SSD}.

Since inclined particles deposit more energy in the SSD, a correction factor -- derived from simulations and a dedicated setup using RPCs to select vertical muons -- is applied to convert omnidirectional (OD) MIP values to vertical equivalents (VE), see right plot in \cref{f:calib_SSD}.
The correction factor estimated with the latter method is
\begin{equation}
 Q^\text{peak}_\text{OD} / Q^\text{peak}_\text{VE} = 1.19 \pm 0.07,
\end{equation}
where the quoted uncertainty of 6\% is a conservative estimate, obtained by adding in quadrature the contributions from PMT instability (4\%) and bar-to-bar light-yield variability (5\%).
Further details on the experimental setup are available in Ref.~\cite{RPC}.

\subsection{Rate Based Calibration. An online estimation of the MIP Peak.}

\begin{figure}
       \centering
       \includegraphics[width=0.5\textwidth]{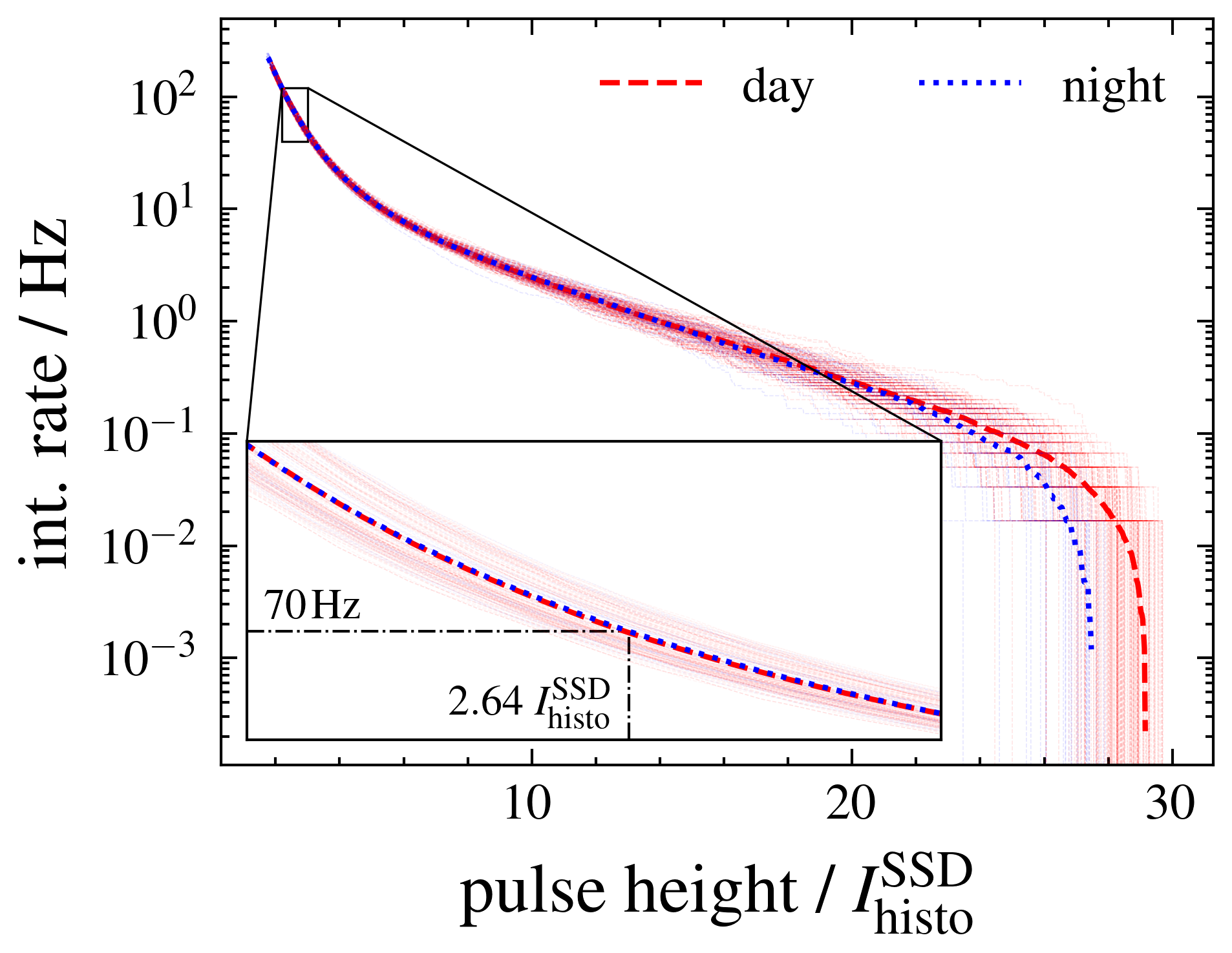}
       \caption{Integral rate of events with a peak height $\geq I_\text{SSD}$.
       The rate remains stable across day/night cycles and different local station hardware.
       A trigger rate of 70\,Hz is achieved for a threshold of $T_{70} = 2.64 \, I_\text{SSD}$.}
       \label{f:calib_online}
\end{figure}

The offline calibration of the SSD detector follows the same model used for the calibration of the PMTs in the WCD, both in terms of the charge distribution (MIP Charge or $Q_\text{SSD}$) and the signal peak (MIP Peak or $I_\text{SSD}$).
While the WCD employs an online procedure primarily to set the detector trigger thresholds, a similar approach is not used for the scintillator.
This is due to hardware limitations at the local station level and because the scintillator operates in slave mode. However, implementing an online estimate would enable real-time monitoring of the detector performance.

To this end, a rate-based model has been developed.
The method employs a single-bin threshold trigger with a variable threshold $T_{70}$, measured in ADC counts.
The number of events $n$ in the SSD that satisfy this trigger condition is counted over an integration window $t_\text{cal}$.
After each $t_\text{cal}$ interval, the threshold is adjusted by a value $\delta$, following the steps described in details in Ref.~\cite{Paul}.
Assuming constant electronic gain, this algorithm guarantees a stable estimate of $T_{70}$.
To relate $T_{70}$ to the MIP peak $I_\text{MIP}$, peak histograms are collected under the condition of a coincident signal in the WCD.
The pulse height spectrum $I$ in the SSD is measured for various stations and times of day.
Using knowledge from these coincidence histograms, the pulse height spectrum is expressed in units of $I_\text{SSD}^\text{histo}$, and the integral event rate is calculated for thresholds of $1$, $2$, etc., times $I_\text{SSD}^\text{histo}$.

This analysis, carried out across all stations and different times of day, reveals no significant dependence on hardware or temperature.
The resulting average rate-threshold relationship is shown in \cref{f:calib_online}.
Notably, a 70\,Hz trigger rate can be achieved by applying a fixed threshold (in units of $I_\text{MIP}$), leading to the rate-based estimator of the MIP peak $I_\text{SSD}^\text{rate} = 2.64\,T_{70}$. It is worth noting that the $70 \text{Hz}$ rate is currently used as a test value; further studies are ongoing to determine the optimal trigger rate for implementing an online estimator of the MIP peak.


\section{Performance}

\begin{figure}
       \centering
       \includegraphics[width=0.7\textwidth]{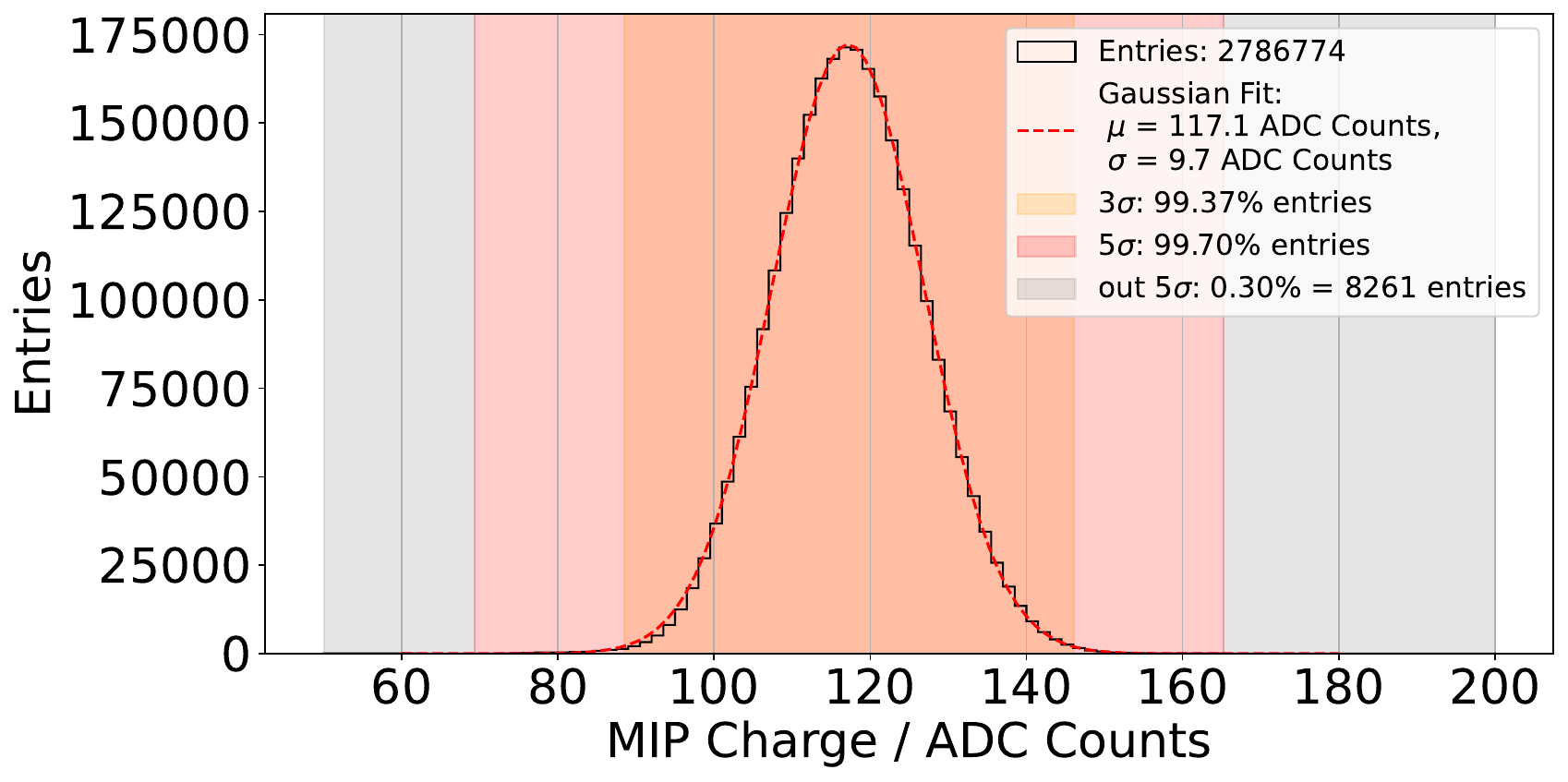}
       \caption{MIP charge distribution for SSD signals collected in Phase-II data.}
       \label{f:mip_distribution}
\end{figure}

\begin{figure}
       \centering
       \includegraphics[width=0.7\textwidth]{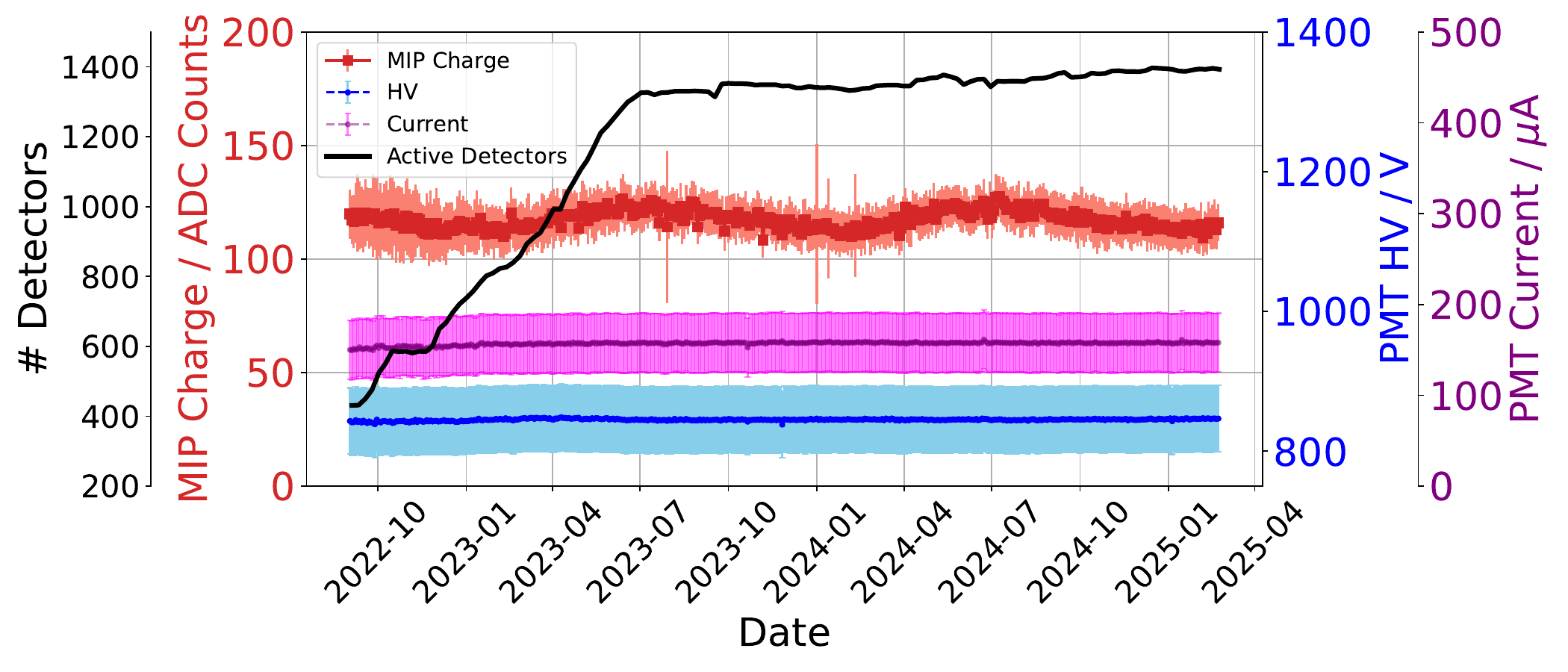}
       \caption{Monitoring data for HV and current for the SSD.
       The averages among all the detectors are showed with relative standard deviation.
       In red the average daily MIP charge distribution.}
       \label{f:monitoring}
\end{figure}

The Surface Scintillator Detector (SSD), part of the AugerPrime upgrade, was fully installed in 2023, with over 1400 units deployed and operational.
This marked a transition phase in the data acquisition of Pierre Auger Observatory, aimed at integrating the new instruments introduced by the upgrade.
Known as Phase\,II, this period includes all data collected with the SSDs since their deployment.
This section evaluates the SSD stability based on operational parameters and early data.

\subsection{MIP Charge Distribution}

The stability of the large number of detectors deployed in the field is essential to ensure a uniform response of the SSD array.
The calibration analysis performed over the first years of data shows that the distribution of the MIP charge is well described by a Gaussian, with mean and standard deviation as shown in \cref{f:mip_distribution}. 
Among all calibrated signals associated with events reconstructed by the surface detector, only 0.30\% fall outside a $5\sigma$ interval from this distribution. 
These outliers are mainly attributed to a small subset of detectors that, for limited periods of time, operated under conditions different from the expected working point. 
In most cases, these deviations are associated with PMTs biased by high voltage values lying outside the typical range of 760 to 980\,V.

In addition to the calibration performance, the monitoring system allows us to evaluate the long-term stability of the detector in terms of high voltage and current, as shown in \cref{f:monitoring}~\cite{Belen}. 
While the MIP charge exhibits seasonal variations -- discussed in the following sections -- the detector demonstrates excellent stability in terms of operating point for the active and functioning units. 
The plot also shows the number of detectors that were active over time, highlighting the progressive development of the array towards its full data-taking capacity in Phase\,II.

\subsection{Seasonal effects}

\begin{figure}
       \centering
       \includegraphics[width=0.7\textwidth]{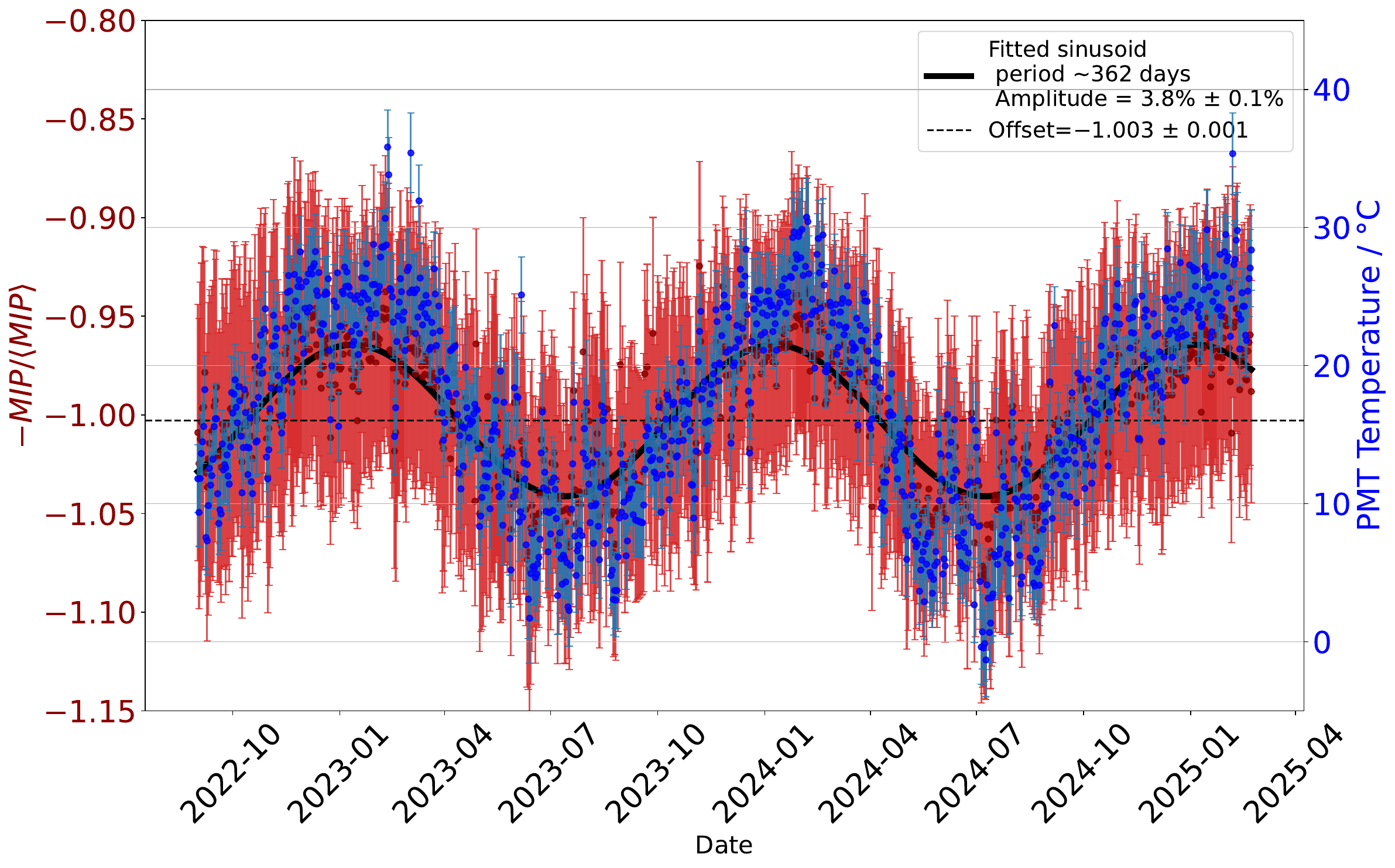}
       \caption{Seasonal variation of the relative MIP charge (red) and the average PMT temperature (blue).}
       \label{f:seasonality}
\end{figure}

The charge distribution of the photomultiplier is affected by temperature, with more stable operation observed at lower temperatures compared to higher ones.
In colder conditions, the signal-to-noise ratio improves, the gain is enhanced, and the MIP charge tends to increase relative to the average level.
This effect, already visible in \cref{f:monitoring}, is further highlighted in \cref{f:seasonality}, where a clear anti-correlation with the average PMT temperature is observed.
To study the behavior of all detectors in the array over time, the MIP charge is normalized as follows:
\begin{itemize}
    \item A dimensionless normalized MIP charge (mip) is calculated for the $k^\text{th}$ detector ($k = 1, \dots,  N_\text{SSD}$) over the full dataset as
    \begin{equation}
    \text{mip}_i^k = \left( \frac{\text{MIP}_i^k}{\text{ADC Counts}} \right) \Bigg/ \left( \frac{1}{N} \sum_{j=1}^{N} \frac{\text{MIP}_j^k}{\text{ADC Counts}} \right)
    \label{relative_mip}
    \end{equation}
    
    \item A daily, array-level relative MIP charge is then defined to track collective variations,
    \begin{equation}
    \frac{\text{MIP}}{\langle \text{MIP} \rangle}(\text{day}) \equiv \frac{1}{\sum_{k=1}^{N_{\text{SSD}}} N_{\text{entries}}^k(\text{day})} 
    \sum_{k=1}^{N_{\text{SSD}}} \sum_{i=1}^{N_{\text{entries}}^k(\text{day})} \text{mip}_i^k
    \label{seasonal_mip}
    \end{equation}
\end{itemize}

The plot in \cref{f:seasonality} shows the evolution of the quantity defined in \cref{seasonal_mip}, inverted in sign to display the oscillation in phase with the temperature.
The dominant oscillation period is found to be approximately 362 days, with an amplitude stable within about 4\% on average. The uncertainty is computed as the standard deviation of the $\text{MIP}/\langle \text{MIP} \rangle$ distribution, and accounts for detector-to-detector fluctuations as well as daily thermal variations, which will be discussed in the next section.

\subsection{Daily temperature correlation}

\begin{figure}
       \centering
       \includegraphics[width=0.7\textwidth]{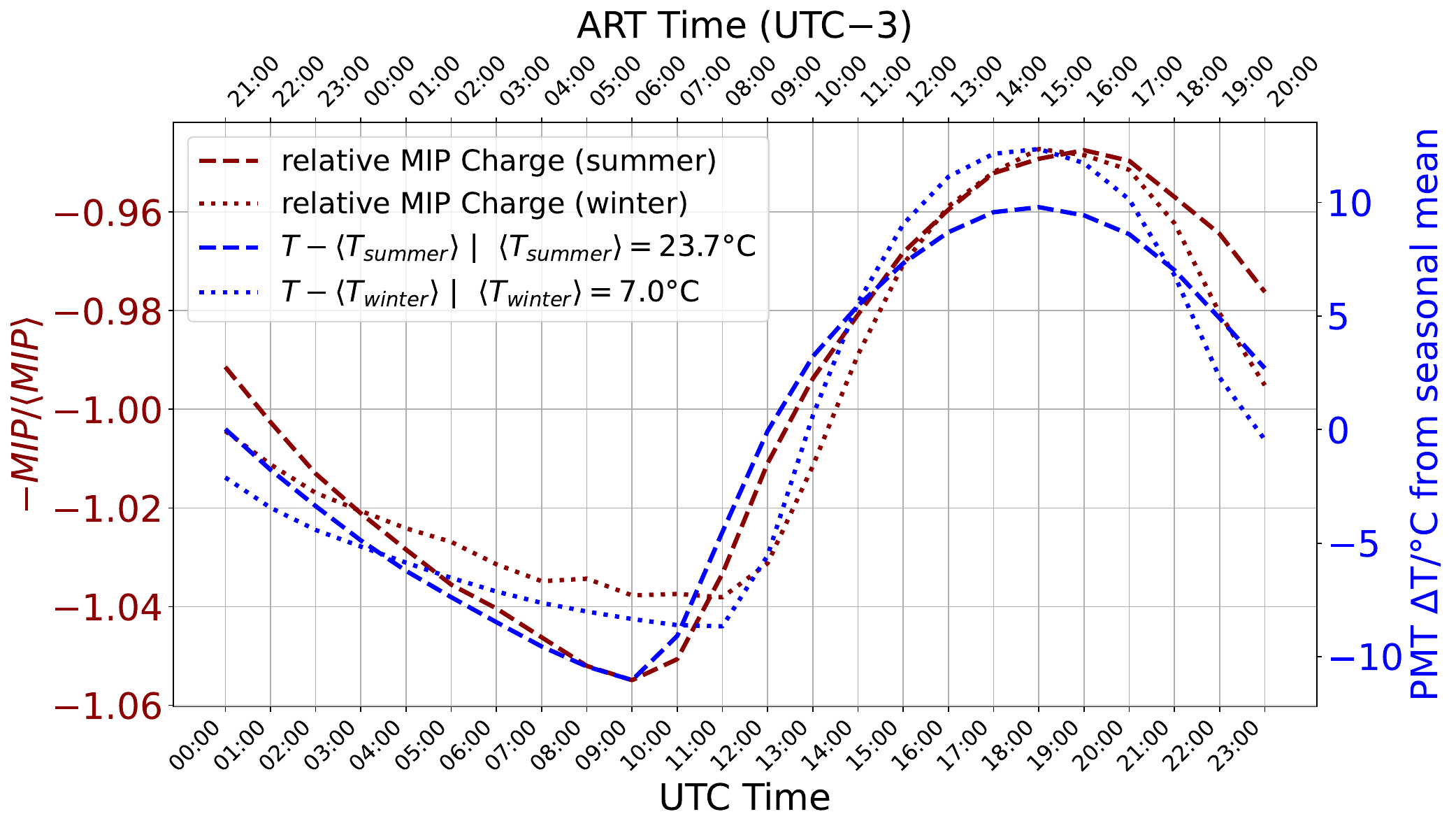}
       \caption{Daily variation of the MIP charge (red) and the average PMT temperature (blue), for two different periods: summer (dashed) and winter (dotted).
       The two horizontal axes display the time in UTC and local Argentinian time, respectively.}
       \label{f:daily_temperature}
\end{figure}

Following the same approach used for the seasonal analysis of the MIP charge fluctuations over time and its correlation with the daily average PMT temperature in the scintillator detectors, this section investigates daily thermal effects by analyzing variations over the course of a day.
Two periods were selected: summer (2024-12-01 to 2025-02-01) and winter (2024-07-01 to 2024-09-01).
The same quantity defined in \cref{seasonal_mip} is now computed hourly instead of daily.

\cref{f:daily_temperature} shows its evolution, plotted with inverted sign to highlight the anti-correlation with PMT temperature.
The left vertical axis indicates the average PMT temperature, expressed as deviation from the seasonal mean ($\langle T_\text{summer}\rangle = 23.7^\circ$C and $\langle T_\text{winter}\rangle = 7.0^\circ$C), to align the curves.
Dashed and dotted lines represent summer and winter data, respectively.

A MIP charge variation of up to 6\% is observed, with a peak-to-peak difference of ${\sim}10\%$, more pronounced in summer.
This suggests greater calibration stability during colder months, when PMTs operate at lower temperatures.

\subsection{Dynamic range and SSD Reconstruction}

\label{dynamic_range}
\begin{figure}
    \centering
    \hspace*{-1.5cm} 
    \begin{subfigure}[b]{0.4\textwidth}
        \includegraphics[height=5.5cm]{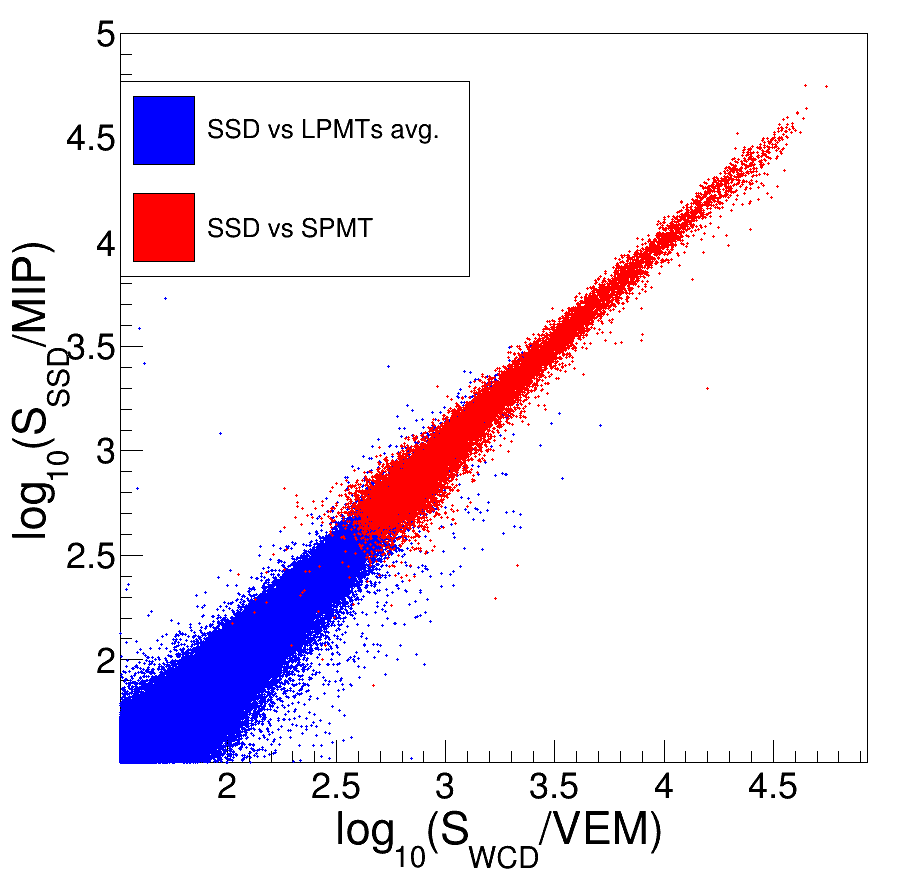}
    \end{subfigure}
    \begin{subfigure}[b]{0.6\textwidth}
        \includegraphics[height=5.5cm]{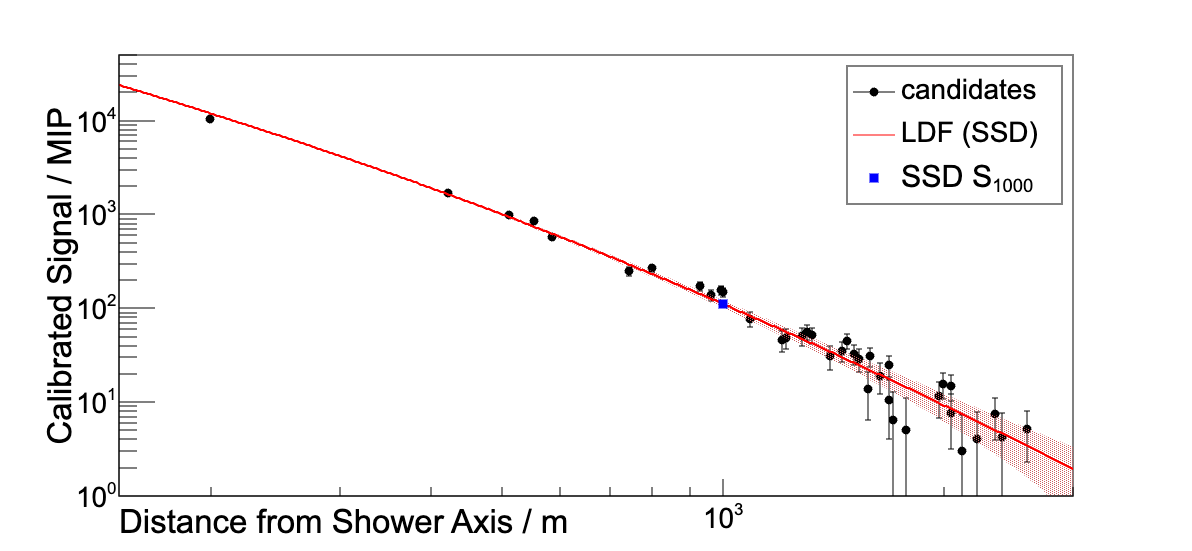}
    \end{subfigure}
    \caption{\textit{Left:} Dynamic range.
    \textit{Right:} Lateral Distribution Function fit of an SD event using the SSD signals.}
    \label{f:reconstruction}
\end{figure}

The upgrade of the surface detector stations has significantly extended the dynamic range, enhancing the ability to accurately measure signals near the shower core and reducing signal saturation.
This issue was especially common in high-energy showers, where the three large PMTs (LPMTs) in the water-Cherenkov detectors often saturated.

To mitigate this, a small PMT (SPMT) was added at the center of each tank, expanding the dynamic range~\cite{Gialex}.
The scintillator detectors were similarly designed to match the SPMT range.
As shown in the left panel of \cref{f:reconstruction}, the Phase-II configuration can handle signals up to several tens of thousands of VEM/MIP before saturation, compared to just a few thousand with LPMTs alone.

Finally, the right panel of \cref{f:reconstruction} shows an event reconstructed using SSD signals.
Station responses are plotted as a function of distance from the shower axis, based on the geometry from standard WCD reconstruction.
The reconstructed event has a primary energy of $E = (49.4 \pm 1.6) \,  \textrm{EeV}$ and a zenith angle of $\theta = (51.73 \pm 0.04)^\circ$.
The data are fitted with a modified NKG function, defined as
\begin{equation}
S(r) = S(r_\text{opt}) \left( \frac{r}{r_\text{opt}}\right)^{\beta} \left( \frac{r + r_\text{s}}{r_\text{opt} + r_\text{s}} \right)^{\beta + \gamma}.
\end{equation}

Since the SSDs are co-located with the WCDs, the same parameters were chosen: $r_\text{opt} = 1000$\,m and $r_\text{s} = 700$\,m, while the parametrization of the shape parameters $\beta$ and $\gamma$ in terms of the shower size and zenith angle is described in Ref.~\cite{Reconstruction}.
As in the case of the WCD, the SSD-based reconstruction allows for the determination of a shower size parameter, $S_{1000}$, defined as the signal at 1000 meters from the shower core, obtained from the fit.

\section{Conclusions}

The installation of over 1400 SSDs as part of the AugerPrime upgrade has significantly enhanced the capabilities of Pierre Auger Observatory.
We presented calibration methods based on atmospheric muons, including a rate-based approach for real-time monitoring.
Performance studies show stable operation across the array, with MIP charge variations well correlated with temperature.
The extended dynamic range enables accurate signal reconstruction even near the shower core.
These results confirm the SSD reliability and its key role in improving mass composition analyses in Phase-II data.

\clearpage
%
%
%


\section*{The Pierre Auger Collaboration}

{\footnotesize\setlength{\baselineskip}{10pt}
\noindent
\begin{wrapfigure}[11]{l}{0.12\linewidth}
\vspace{-4pt}
\includegraphics[width=0.98\linewidth]{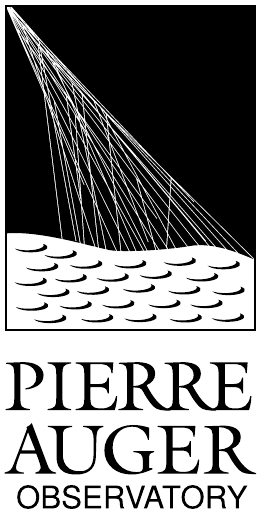}
\end{wrapfigure}
\begin{sloppypar}\noindent
\input{latex_authorlist_authors}
\end{sloppypar}
\begin{center}
\end{center}

\vspace{1ex}
\input{latex_authorlist_institutions}

\input{acknowledgments}
}

\end{document}

%% file: latex_authorlist_authors.tex
A.~Abdul Halim$^{13}$,
P.~Abreu$^{70}$,
M.~Aglietta$^{53,51}$,
I.~Allekotte$^{1}$,
K.~Almeida Cheminant$^{78,77}$,
A.~Almela$^{7,12}$,
R.~Aloisio$^{44,45}$,
J.~Alvarez-Mu\~niz$^{76}$,
A.~Ambrosone$^{44}$,
J.~Ammerman Yebra$^{76}$
% ...

%% file: latex_authorlist_institutions.tex
\begin{description}[labelsep=0.2em,align=right,labelwidth=0.7em,labelindent=0em,leftmargin=2em,noitemsep,before={\renewcommand\makelabel[1]{##1 }}]
\item[$^{1}$] Centro At\'omico Bariloche and Instituto Balseiro (CNEA-UNCuyo-CONICET), San Carlos de Bariloche, Argentina
\item[$^{2}$] Departamento de F\'\i{}sica and Departamento de Ciencias de la Atm\'osfera y los Oc\'eanos, FCEyN, Universidad de Buenos Aires and CONICET, Buenos Aires, Argentina
\item[$^{3}$] IFLP, Universidad Nacional de La Plata and CONICET, La Plata, Argentina
\item[$^{4}$] Instituto de Astronom\'\i{}a y F\'\i{}sica del Espacio (IAFE, CONICET-UBA), Buenos Aires, Argentina
\item[$^{5}$] Instituto de F\'\i{}sica de Rosario (IFIR) -- CONICET/U.N.R.\ and Facultad de Ciencias Bioqu\'\i{}micas y Farmac\'euticas U.N.R., Rosario, Argentina
\item[$^{6}$] Instituto de Tecnolog\'\i{}as en Detecci\'on y Astropart\'\i{}culas (CNEA, CONICET, UNSAM), and Universidad Tecnol\'ogica Nacional -- Facultad Regional Mendoza (CONICET/CNEA), Mendoza, Argentina
\item[$^{7}$] Instituto de Tecnolog\'\i{}as en Detecci\'on y Astropart\'\i{}culas (CNEA, CONICET, UNSAM), Buenos Aires, Argentina
\item[$^{8}$] International Center of Advanced Studies and Instituto de Ciencias F\'\i{}sicas, ECyT-UNSAM and CONICET, Campus Miguelete -- San Mart\'\i{}n, Buenos Aires, Argentina
\item[$^{9}$] Laboratorio Atm\'osfera -- Departamento de Investigaciones en L\'aseres y sus Aplicaciones -- UNIDEF (CITEDEF-CONICET), Argentina
\item[$^{10}$] Observatorio Pierre Auger, Malarg\"ue, Argentina
\item[$^{11}$] Observatorio Pierre Auger and Comisi\'on Nacional de Energ\'\i{}a At\'omica, Malarg\"ue, Argentina
\item[$^{12}$] Universidad Tecnol\'ogica Nacional -- Facultad Regional Buenos Aires, Buenos Aires, Argentina
\item[$^{13}$] University of Adelaide, Adelaide, S.A., Australia
\item[$^{14}$] Universit\'e Libre de Bruxelles (ULB), Brussels, Belgium
\item[$^{15}$] Vrije Universiteit Brussels, Brussels, Belgium
\item[$^{16}$] Centro Brasileiro de Pesquisas Fisicas, Rio de Janeiro, RJ, Brazil
\item[$^{17}$] Centro Federal de Educa\c{c}\~ao Tecnol\'ogica Celso Suckow da Fonseca, Petropolis, Brazil
\item[$^{18}$] Instituto Federal de Educa\c{c}\~ao, Ci\^encia e Tecnologia do Rio de Janeiro (IFRJ), Brazil
\item[$^{19}$] Universidade de S\~ao Paulo, Escola de Engenharia de Lorena, Lorena, SP, Brazil
\item[$^{20}$] Universidade de S\~ao Paulo, Instituto de F\'\i{}sica de S\~ao Carlos, S\~ao Carlos, SP, Brazil
\item[$^{21}$] Universidade de S\~ao Paulo, Instituto de F\'\i{}sica, S\~ao Paulo, SP, Brazil
\item[$^{22}$] Universidade Estadual de Campinas (UNICAMP), IFGW, Campinas, SP, Brazil
\item[$^{23}$] Universidade Estadual de Feira de Santana, Feira de Santana, Brazil
\item[$^{24}$] Universidade Federal de Campina Grande, Centro de Ciencias e Tecnologia, Campina Grande, Brazil
\item[$^{25}$] Universidade Federal do ABC, Santo Andr\'e, SP, Brazil
\item[$^{26}$] Universidade Federal do Paran\'a, Setor Palotina, Palotina, Brazil
\item[$^{27}$] Universidade Federal do Rio de Janeiro, Instituto de F\'\i{}sica, Rio de Janeiro, RJ, Brazil
\item[$^{28}$] Universidad de Medell\'\i{}n, Medell\'\i{}n, Colombia
\item[$^{29}$] Universidad Industrial de Santander, Bucaramanga, Colombia
\item[$^{30}$] Charles University, Faculty of Mathematics and Physics, Institute of Particle and Nuclear Physics, Prague, Czech Republic
\item[$^{31}$] Institute of Physics of the Czech Academy of Sciences, Prague, Czech Republic
\item[$^{32}$] Palacky University, Olomouc, Czech Republic
\item[$^{33}$] CNRS/IN2P3, IJCLab, Universit\'e Paris-Saclay, Orsay, France
\item[$^{34}$] Laboratoire de Physique Nucl\'eaire et de Hautes Energies (LPNHE), Sorbonne Universit\'e, Universit\'e de Paris, CNRS-IN2P3, Paris, France
\item[$^{35}$] Univ.\ Grenoble Alpes, CNRS, Grenoble Institute of Engineering Univ.\ Grenoble Alpes, LPSC-IN2P3, 38000 Grenoble, France
\item[$^{36}$] Universit\'e Paris-Saclay, CNRS/IN2P3, IJCLab, Orsay, France
\item[$^{37}$] Bergische Universit\"at Wuppertal, Department of Physics, Wuppertal, Germany
\item[$^{38}$] Karlsruhe Institute of Technology (KIT), Institute for Experimental Particle Physics, Karlsruhe, Germany
\item[$^{39}$] Karlsruhe Institute of Technology (KIT), Institut f\"ur Prozessdatenverarbeitung und Elektronik, Karlsruhe, Germany
\item[$^{40}$] Karlsruhe Institute of Technology (KIT), Institute for Astroparticle Physics, Karlsruhe, Germany
\item[$^{41}$] RWTH Aachen University, III.\ Physikalisches Institut A, Aachen, Germany
\item[$^{42}$] Universit\"at Hamburg, II.\ Institut f\"ur Theoretische Physik, Hamburg, Germany
\item[$^{43}$] Universit\"at Siegen, Department Physik -- Experimentelle Teilchenphysik, Siegen, Germany
\item[$^{44}$] Gran Sasso Science Institute, L'Aquila, Italy
\item[$^{45}$] INFN Laboratori Nazionali del Gran Sasso, Assergi (L'Aquila), Italy
\item[$^{46}$] INFN, Sezione di Catania, Catania, Italy
\item[$^{47}$] INFN, Sezione di Lecce, Lecce, Italy
\item[$^{48}$] INFN, Sezione di Milano, Milano, Italy
\item[$^{49}$] INFN, Sezione di Napoli, Napoli, Italy
\item[$^{50}$] INFN, Sezione di Roma ``Tor Vergata'', Roma, Italy
\item[$^{51}$] INFN, Sezione di Torino, Torino, Italy
\item[$^{52}$] Istituto di Astrofisica Spaziale e Fisica Cosmica di Palermo (INAF), Palermo, Italy
\item[$^{53}$] Osservatorio Astrofisico di Torino (INAF), Torino, Italy
\item[$^{54}$] Politecnico di Milano, Dipartimento di Scienze e Tecnologie Aerospaziali , Milano, Italy
\item[$^{55}$] Universit\`a del Salento, Dipartimento di Matematica e Fisica ``E.\ De Giorgi'', Lecce, Italy
\item[$^{56}$] Universit\`a dell'Aquila, Dipartimento di Scienze Fisiche e Chimiche, L'Aquila, Italy
\item[$^{57}$] Universit\`a di Catania, Dipartimento di Fisica e Astronomia ``Ettore Majorana``, Catania, Italy
\item[$^{58}$] Universit\`a di Milano, Dipartimento di Fisica, Milano, Italy
\item[$^{59}$] Universit\`a di Napoli ``Federico II'', Dipartimento di Fisica ``Ettore Pancini'', Napoli, Italy
\item[$^{60}$] Universit\`a di Palermo, Dipartimento di Fisica e Chimica ''E.\ Segr\`e'', Palermo, Italy
\item[$^{61}$] Universit\`a di Roma ``Tor Vergata'', Dipartimento di Fisica, Roma, Italy
\item[$^{62}$] Universit\`a Torino, Dipartimento di Fisica, Torino, Italy
\item[$^{63}$] Benem\'erita Universidad Aut\'onoma de Puebla, Puebla, M\'exico
\item[$^{64}$] Unidad Profesional Interdisciplinaria en Ingenier\'\i{}a y Tecnolog\'\i{}as Avanzadas del Instituto Polit\'ecnico Nacional (UPIITA-IPN), M\'exico, D.F., M\'exico
\item[$^{65}$] Universidad Aut\'onoma de Chiapas, Tuxtla Guti\'errez, Chiapas, M\'exico
\item[$^{66}$] Universidad Michoacana de San Nicol\'as de Hidalgo, Morelia, Michoac\'an, M\'exico
\item[$^{67}$] Universidad Nacional Aut\'onoma de M\'exico, M\'exico, D.F., M\'exico
\item[$^{68}$] Institute of Nuclear Physics PAN, Krakow, Poland
\item[$^{69}$] University of \L{}\'od\'z, Faculty of High-Energy Astrophysics,\L{}\'od\'z, Poland
\item[$^{70}$] Laborat\'orio de Instrumenta\c{c}\~ao e F\'\i{}sica Experimental de Part\'\i{}culas -- LIP and Instituto Superior T\'ecnico -- IST, Universidade de Lisboa -- UL, Lisboa, Portugal
\item[$^{71}$] ``Horia Hulubei'' National Institute for Physics and Nuclear Engineering, Bucharest-Magurele, Romania
\item[$^{72}$] Institute of Space Science, Bucharest-Magurele, Romania
\item[$^{73}$] Center for Astrophysics and Cosmology (CAC), University of Nova Gorica, Nova Gorica, Slovenia
\item[$^{74}$] Experimental Particle Physics Department, J.\ Stefan Institute, Ljubljana, Slovenia
\item[$^{75}$] Universidad de Granada and C.A.F.P.E., Granada, Spain
\item[$^{76}$] Instituto Galego de F\'\i{}sica de Altas Enerx\'\i{}as (IGFAE), Universidade de Santiago de Compostela, Santiago de Compostela, Spain
\item[$^{77}$] IMAPP, Radboud University Nijmegen, Nijmegen, The Netherlands
\item[$^{78}$] Nationaal Instituut voor Kernfysica en Hoge Energie Fysica (NIKHEF), Science Park, Amsterdam, The Netherlands
\item[$^{79}$] Stichting Astronomisch Onderzoek in Nederland (ASTRON), Dwingeloo, The Netherlands
\item[$^{80}$] Universiteit van Amsterdam, Faculty of Science, Amsterdam, The Netherlands
\item[$^{81}$] Case Western Reserve University, Cleveland, OH, USA
\item[$^{82}$] Colorado School of Mines, Golden, CO, USA
\item[$^{83}$] Department of Physics and Astronomy, Lehman College, City University of New York, Bronx, NY, USA
\item[$^{84}$] Michigan Technological University, Houghton, MI, USA
\item[$^{85}$] New York University, New York, NY, USA
\item[$^{86}$] University of Chicago, Enrico Fermi Institute, Chicago, IL, USA
\item[$^{87}$] University of Delaware, Department of Physics and Astronomy, Bartol Research Institute, Newark, DE, USA
\item[] -----
\item[$^{a}$] Max-Planck-Institut f\"ur Radioastronomie, Bonn, Germany
\item[$^{b}$] also at Kapteyn Institute, University of Groningen, Groningen, The Netherlands
\item[$^{c}$] School of Physics and Astronomy, University of Leeds, Leeds, United Kingdom
\item[$^{d}$] Fermi National Accelerator Laboratory, Fermilab, Batavia, IL, USA
\item[$^{e}$] Pennsylvania State University, University Park, PA, USA
\item[$^{f}$] Colorado State University, Fort Collins, CO, USA
\item[$^{g}$] Louisiana State University, Baton Rouge, LA, USA
\item[$^{h}$] now at Graduate School of Science, Osaka Metropolitan University, Osaka, Japan
\item[$^{i}$] Institut universitaire de France (IUF), France
\item[$^{j}$] now at Technische Universit\"at Dortmund and Ruhr-Universit\"at Bochum, Dortmund and Bochum, Germany
\end{description}

%% file: acknowledgments.tex
\section*{Acknowledgments}

\begin{sloppypar}
The successful installation, commissioning, and operation of the Pierre
Auger Observatory would not have been possible without the strong
commitment and effort from the technical and administrative staff in
Malarg\"ue. We are very grateful to the following agencies and
organizations for financial support:
\end{sloppypar}

\begin{sloppypar}
Argentina -- Comisi\'on Nacional de Energ\'\i{}a At\'omica; Agencia Nacional de
Promoci\'on Cient\'\i{}fica y Tecnol\'ogica (ANPCyT); Consejo Nacional de
Investigaciones Cient\'\i{}ficas y T\'ecnicas (CONICET); Gobierno de la
Provincia de Mendoza; Municipalidad de Malarg\"ue; NDM Holdings and Valle
Las Le\~nas; in gratitude for their continuing cooperation over land
access; Australia -- the Australian Research Council; Belgium -- Fonds
de la Recherche Scientifique (FNRS); Research Foundation Flanders (FWO),
Marie Curie Action of the European Union Grant No.~101107047; Brazil --
Conselho Nacional de Desenvolvimento Cient\'\i{}fico e Tecnol\'ogico (CNPq);
Financiadora de Estudos e Projetos (FINEP); Funda\c{c}\~ao de Amparo \`a
Pesquisa do Estado de Rio de Janeiro (FAPERJ); S\~ao Paulo Research
Foundation (FAPESP) Grants No.~2019/10151-2, No.~2010/07359-6 and
No.~1999/05404-3; Minist\'erio da Ci\^encia, Tecnologia, Inova\c{c}\~oes e
Comunica\c{c}\~oes (MCTIC); Czech Republic -- GACR 24-13049S, CAS LQ100102401,
MEYS LM2023032, CZ.02.1.01/0.0/0.0/16{\textunderscore}013/0001402,
CZ.02.1.01/0.0/0.0/18{\textunderscore}046/0016010 and
CZ.02.1.01/0.0/0.0/17{\textunderscore}049/0008422 and CZ.02.01.01/00/22{\textunderscore}008/0004632;
France -- Centre de Calcul IN2P3/CNRS; Centre National de la Recherche
Scientifique (CNRS); Conseil R\'egional Ile-de-France; D\'epartement
Physique Nucl\'eaire et Corpusculaire (PNC-IN2P3/CNRS); D\'epartement
Sciences de l'Univers (SDU-INSU/CNRS); Institut Lagrange de Paris (ILP)
Grant No.~LABEX ANR-10-LABX-63 within the Investissements d'Avenir
Programme Grant No.~ANR-11-IDEX-0004-02; Germany -- Bundesministerium
f\"ur Bildung und Forschung (BMBF); Deutsche Forschungsgemeinschaft (DFG);
Finanzministerium Baden-W\"urttemberg; Helmholtz Alliance for
Astroparticle Physics (HAP); Helmholtz-Gemeinschaft Deutscher
Forschungszentren (HGF); Ministerium f\"ur Kultur und Wissenschaft des
Landes Nordrhein-Westfalen; Ministerium f\"ur Wissenschaft, Forschung und
Kunst des Landes Baden-W\"urttemberg; Italy -- Istituto Nazionale di
Fisica Nucleare (INFN); Istituto Nazionale di Astrofisica (INAF);
Ministero dell'Universit\`a e della Ricerca (MUR); CETEMPS Center of
Excellence; Ministero degli Affari Esteri (MAE), ICSC Centro Nazionale
di Ricerca in High Performance Computing, Big Data and Quantum
Computing, funded by European Union NextGenerationEU, reference code
CN{\textunderscore}00000013; M\'exico -- Consejo Nacional de Ciencia y Tecnolog\'\i{}a
(CONACYT) No.~167733; Universidad Nacional Aut\'onoma de M\'exico (UNAM);
PAPIIT DGAPA-UNAM; The Netherlands -- Ministry of Education, Culture and
Science; Netherlands Organisation for Scientific Research (NWO); Dutch
national e-infrastructure with the support of SURF Cooperative; Poland
-- Ministry of Education and Science, grants No.~DIR/WK/2018/11 and
2022/WK/12; National Science Centre, grants No.~2016/22/M/ST9/00198,
2016/23/B/ST9/01635, 2020/39/B/ST9/01398, and 2022/45/B/ST9/02163;
Portugal -- Portuguese national funds and FEDER funds within Programa
Operacional Factores de Competitividade through Funda\c{c}\~ao para a Ci\^encia
e a Tecnologia (COMPETE); Romania -- Ministry of Research, Innovation
and Digitization, CNCS-UEFISCDI, contract no.~30N/2023 under Romanian
National Core Program LAPLAS VII, grant no.~PN 23 21 01 02 and project
number PN-III-P1-1.1-TE-2021-0924/TE57/2022, within PNCDI III; Slovenia
-- Slovenian Research Agency, grants P1-0031, P1-0385, I0-0033, N1-0111;
Spain -- Ministerio de Ciencia e Innovaci\'on/Agencia Estatal de
Investigaci\'on (PID2019-105544GB-I00, PID2022-140510NB-I00 and
RYC2019-027017-I), Xunta de Galicia (CIGUS Network of Research Centers,
Consolidaci\'on 2021 GRC GI-2033, ED431C-2021/22 and ED431F-2022/15),
Junta de Andaluc\'\i{}a (SOMM17/6104/UGR and P18-FR-4314), and the European
Union (Marie Sklodowska-Curie 101065027 and ERDF); USA -- Department of
Energy, Contracts No.~DE-AC02-07CH11359, No.~DE-FR02-04ER41300,
No.~DE-FG02-99ER41107 and No.~DE-SC0011689; National Science Foundation,
Grant No.~0450696, and NSF-2013199; The Grainger Foundation; Marie
Curie-IRSES/EPLANET; European Particle Physics Latin American Network;
and UNESCO.
\end{sloppypar}

%% file: main.bbl
\begingroup
\footnotesize
\begin{thebibliography}{99}

\bibitem{SSD} A.~Abdul Halim \emph{et al.} [Pierre Auger Coll.], arXiv:2507.07762.
\bibitem{Fabio} F.~Convenga \emph{et al.} [Pierre Auger Coll.], PoS(\textbf{ICRC2023})392.
\bibitem{RPC} A.~Aab \emph{et al.} [Pierre Auger Coll.], JINST \textbf{15} (2020) P09002.
\bibitem{Paul} P.~Filip \emph{et al.} [Pierre Auger Coll.], PoS(\textbf{UHECR2024})085.
\bibitem{Belen} B.~Andrada \emph{et al.} [Pierre Auger Coll.], PoS(\textbf{ICRC2025})176.
\bibitem{Gialex} G.A.~Anastasi \emph{et al.} [Pierre Auger Coll.], PoS(\textbf{ICRC2023})343.
\bibitem{Reconstruction} A.~Aab \emph{et al.} [Pierre Auger Coll.], JINST \textbf{15} (2020) P10021.

\end{thebibliography}
\endgroup
